# BIOMIMETIC SPACE-VARIANT SAMPLING IN A VISION PROSTHESIS IMPROVES THE USER'S SKILL IN A LOCALIZATION TASK

B. Durette[1,2], L. Gamond[1], S. Hanneton[3], D. Alleysson[2], J. Hérault[1]

[1] Department of Images and Signals, Gipsa-Lab, CNRS UMR 5216, UJF, INPG, Grenoble, France
[2] Laboratory of Psychology et Neuro-cognition, CNRS UMR 5105, UPMF, Grenoble, France
[3] Laboratory of Neurophysics and Physiology, CNRS UMR 8119, Université Paris V, Paris, France

**Abstract:** In this experiment, we test the hypothesis of whether a 'retina-like' space variant sampling pattern can improve the efficiency of a visual prosthesis. Subjects wearing a visuo-auditory substitution system were tested for their ability to point at visual targets. The test group (space-variant sampling), performed significantly better than the control group (uniform sampling). The pointing accuracy was enhanced, as was the speed to find the target. Surprisingly, the time spanned to complete the training was also reduced, suggesting that this space-variant sampling scheme facilitates the mastering of sensorimotor contingencies.

**Keywords**: Visual prosthesis, space-variant sampling, design principles, sensori-motricity, learning.

## 1. Introduction

Visual prosthesis are devices that interface a video-camera with the brain at different levels : either directly implanted on the retina or on the cortex surface (for a review, see e.g. Zrenner, 2002, Margalit *et al.*, 2002), or by means of a sane substitute sense, most of the time the tactile sense (Sampaio *et al.*, 2001, Bach-y-Rita *et al.*, 2004, Kajimoto *et al.*, 2006), or the auditory sense (Meijer, 1992, Auvray et al, 2005). Those latest devices are called « sensory substitution systems ».

The main difference between visual prosthesis and natural vision systems is likely to be the number of stimulation points available. As compared to the 6 million cones in the human eye or the 80000 pixels of a video camera, visual prosthesis resolution, all categories considered, ranges from 64 (cortical implant, Dobelle *et al.*, 2000) to 896 synchronous stimulation points (VideoTact tactile array, ForeThought Dev.). Wider arrays are under development, however their spatial resolution will probably be limited soon, not because of technology, but by the sensitive substrate itself (Zrenner, 2002). The gap between natural vision systems and interfacing solutions makes the question of resolution reduction a critical point for vision prosthesis. Most of the time, the resolution reduction is done by a uniform subsampling, either directly on the picture (Bach-y-Rita *et al.*, 1969, Meijer, 1992, Thompson *et al.*, 2003), or after a preliminary signal processing stage like uniform averaging (Sampaio *et al.*, 2001, Harvey & Sawan, 1996) or edge detection (Dobelle, 2000, Kajimoto *et al.*, 2006). However, when applied to large fields of vision, uniform subsampling leads to a low global resolution.

To answer this problem, natural systems have adopted a space-variant sampling principle. The visual system of primates, for instance, possesses a highly sampled "foveal" region, at the center of the visual field (about 3° wide). Sampling distribution then rapidly decreases with eccentricity (Osterberg, 1935). This feature is generally understood as a focus/context strategy, the visual system being able to roughly detect an object of interest in its field of vision and then to direct his fovea to it for identification if necessary. Space-variant sampling has often been mentioned as a possible tool to enhance visual prosthesis (e.g. Eckmiller *et al.*, 2005, Naghdy, 2006). To our knowledge, it has been implemented in only two devices: the PSVA (Capelle *et al.*, 1998) and the VAS (Gonzalez-Mora, 2003). However, the sampling distributions were determined empirically, and no comparison was made with othe possible distributions, like uniform.

In this article, we show how recent advances into the comprehension of visual perception in terms of sensorimotor contingencies (O'Regan & Noë, 2002) as well as knowledge of signal processing in the primate early visual system (Hérault & Durette, 2007) give new arguments and new tools to address the question of space-variant sampling in visual prosthesis. We then propose a particular sampling distribution and test the hypothesis of whether it can improve the efficiency of a visual prosthesis. Blindfolded subjects wearing a visuo-auditory substitution system (TheVIBE, Auvray *et al.* 2005) were tested for their ability to point at visual targets. The test group (space-variant sampling) performed significantly better than the control group (uniform sampling). The pointing accuracy was enhanced, as was the speed to find the target. Surprisingly, the time spanned to complete the training was also reduced, suggesting that this space-variant sampling may facilitate the mastering of sensorimotor contingencies.

## 2. Theoretical Aspects: Sampling, Sensorimotricity and Vision

### 2.1. A biomimetic inspiration

To address the question of space-variant sampling, we first needed to choose among all possible sampling distribution laws. The law we choose is directly inspired from observation on primates visual system. As described by Schwartz *et al.* (1980), mapping of the visual world onto the primate's primary visual cortex is highly space-variant. In particular, the inverse ratio between a distance in the visual world and its correlate in its cortical projection, often referred to as the "cortical magnification factor", strongly decreases with eccentricity. From neurophysiological observations, Schwartz *et al.* derived a "global retinotopic mapping" of the visual information to the visual cortex described as the complex logarithm of a linear function of eccentricity. With z being a point in the visual plane and w it's projection on the cortical plane, transformation between z and w can be written $w = \log(z+a)$, a being a constant. With $z = \rho e^{i\theta}$ and $w = \rho' e^{i\theta'}$, this formula links the eccentricity $\rho$ of a point in the visual world to it's "eccentricity" $\rho'$ in the visual cortex:

$$\rho' = \log(\rho + a)$$

### 2.2. Sensori-motor arguments to a logarithmic mapping

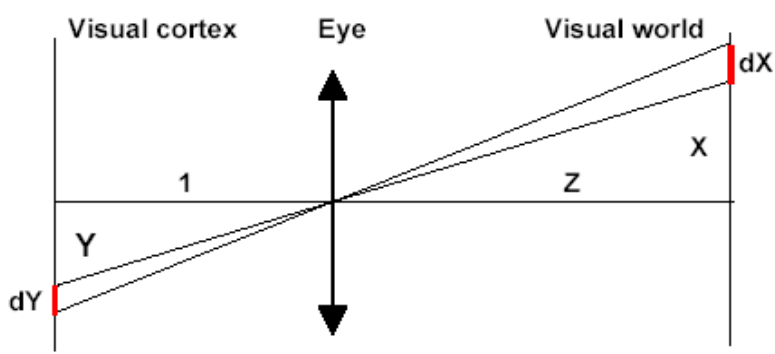

Figure 1: Simplified illustration of the retino-cortical mapping in two dimensions

Biomimeticity and focus/context strategy are not the only arguments in favor of a logarithmic mapping of the visual eccentricity: it may also bring new regularities in sensori-motor coupling. Let us compare the properties of a logarithmic cortical mapping with respect to a linear one. An object of size dX is positioned at the eccentricity X in a visual plane at a distance Z from the observer (fig. 1). It's correlate is an object of size dY at the eccentricity Y in the cortex. The central lens symbolizes the eye. For simplification purpose, the distance between the central lens and the cortex plane is chosen as unit distance. With linear mapping, we obtain (c index stands for logarithmic coordinates):

B. Durette, L. Gamond, S. Hanneton, D. Alleysson, J. Hérault

$Y = \dfrac{X}{Z}$ (1) which, by differentiation, gives $dY = \dfrac{X}{Z}\left(\dfrac{dX}{X} - \dfrac{dZ}{Z}\right)$ (2)

Assuming a logarithmic cortical mapping, eq. 1 becomes

$Y_c = \log\left(\dfrac{X}{Z}\right)$ (3) which, by differentiation, gives $dY_c = \left(\dfrac{dX}{X} - \dfrac{dZ}{Z}\right)$ (4),

Formula (4) brings new regularities in the link between the visual world and its cortical correlate with respect to sensori-motor coupling, particularly in the case of a motion along the direction Z:

1- An object O contained in the vertical plane (dZ=0) has a constant size on the cortex regardless of the viewing distance.

Indeed, dZ=0 in eq 3 gives $dY_c = \left(\dfrac{dX}{X}\right)$ (5). As dX and X are constant, dYc is also constant.

2– When approaching at a constant velocity v toward an object O contained in the vertical plane (dX/dt=0), the velocity of its projection on the cortical plane is inversely proportionnal to the *time to contact (T)* between the object and the observer.

Indeed, temporal derivation of eq. 3 with dX/dt=0 gives $\left|\dfrac{dYc}{dt}\right| = \left|-\dfrac{1}{Z}\dfrac{dZ}{dt}\right| = \dfrac{1}{T}$ (6).

Thus, logarithmic mapping of the visual world simplifies the relationship between the subject's motion and the changes it implies in his sensations. It is the reason why we claim that it brings new sensori-motor regularities. Our hypothesis is that the use of such a mapping in a visual prosthesis should enhance it's efficiency.

### 2.3. Implementation of the logarithmic mapping

The mapping function we use is a Michaelis-Menten law which is linear for small eccentricities, logarithmic for medium ones and then saturates. One of its major advantages with respect to a logarithmic law is that it is bounded, thus fitting with the finite cortical space. With $\rho$ being the eccentricity of a point in the visual space and $\rho'$ its correlate in the cortical space, this law can be written :

$\rho' = \rho'_{\lim}\dfrac{\rho}{\rho + \rho_o}$ (7)

$\rho'_{lim}$ and $\rho_0$ are determined so that the image size is preserved and that the central region magnification factor (lately referred to as R0) is adjustable. The central magnification factor R0 is define as the ratio between sampling density at the excentricity $\rho$=0 over global sampling density.

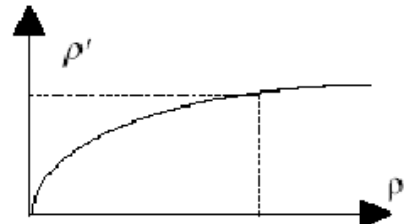

Figure 2 : Mapping of the cortical eccentricity $\rho'$ as a function of the visual eccentricity $\rho$.

### 3. Apparatus: Building Space-Variant Retinas for TheVIBE

### 3.1. TheVIBE auditory substitution system

TheVibe device is an experimental system for the conversion of images into sound patterns (Auvray *et al.* 2005). The image is sampled by a set of "receptive fields". Each receptive field is a cluster of random localized pixels. Those receptive fields compose TheVibe's virtual retina (fig 3). The auditory output is composed of a sum of sinusoidal sounds produced by virtual sound "sources," each corresponding to one of the retina's "receptive fields." The frequency and the inter-aural disparity of each sources are determined by the co-ordinates of the receptive field's pixels in the image (Fig 3, squares).The sound's amplitude is determined by the mean luminosity of the pixels of the corresponding receptive field (Fig. 3, crosses). The ability to freely define the receptive field's position and configuration makes TheVibe a particularly proper tool to address the question of space variant mapping.

B. Durette, L. Gamond, S. Hanneton, D. Alleysson, J. Hérault

*3.2. Uniform and space-variant retina design*

To design standard uniform retinas for TheVibe, we cut the 320 x 240 image into a set of 16 x 12 cells, each side 20x20 pixels. A receptive field origin is chosen at a random position in each cells. Each receptive field is composed of 10 sampling pixels chosen randomly in a 20 x 20 box centered on the receptive field origin so that overlap with other receptive field is possible. Our retinas were composed by 192 receptive fields. Frequency and inter-aural disparity followed a linear mapping of the vertical (resp. horizontal) position. Frequencies ranged from 300 to 3000 Hz. To design space-variant retinas, a logarithmic mapping as defined in section 2, with a central magnification factor R0=2, was applied to uniform retinas. The result is illustrated in fig 3. Initial linear mapping in the auditory space was kept.

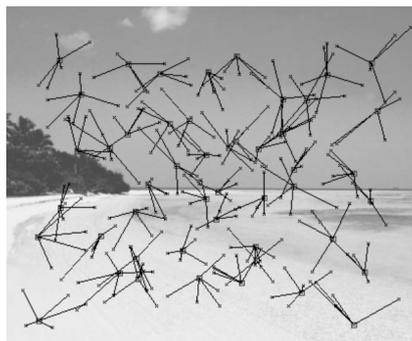

Fig 3 : A space-variant logarithmic (R0=2) retina for TheVibe. Note that the extent of the receptive fields increases with eccentricity. In particular, our mapping preserves overlap.

**4. Performance Assessment: A 'Contact' Task**

The test protocol was inspired by the work of Auvray (2004) which study immersion stages in a sensory substitution system. Our task aims at testing the first stage of immersion, i.e. the 'contact' stage, where the user learns sensorimotor rules to stabilize the stimulus and maintain contact with it. We extended it to the ability to direct the camera toward the target in a stable configuration, i.e. to consistently place a visual target at a systematical location in the visual field. This place was not necessarily the objective center of vision, which was never mentioned in the experiment. The task was thus totally non-supervized.

*4.1. Experimental setup*

The subject was placed in front of a screen were a white target, 8° in diameter was presented on a black background at different locations (fig. 4). The projected picture covers exactly the 78° x 58° field of view of the substitution device. Targets were generated with PsychToolbox (Brainard, 1997). The experiment was separated in three stages divided by 5 mn breaks.

The first stage was "free exploration". In the first 5 minutes, the subject, standing in front of a unique target, freely explored it's visual field. He was allowed to move his head and body as he wished. The next 5 minutes, subject was asked to move to the right and to the left while keeping the target fixed with respect to the vision device, first at ½ m of the screen and then at 1m.

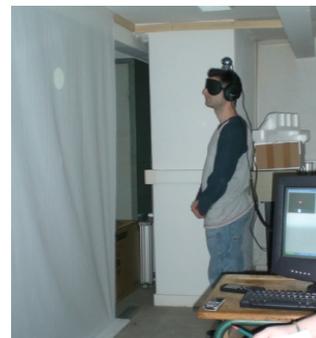

Fig 4: Experimental room and setup

The second stage was "masking". In this task, the subject was seated in front of the screen. For the first thirty randomized target positions, he had to point to the target with his head and then to mask it with his hand, keeping his arm extended (so he does not mask the whole camera aperture). For the next thirty trials, he had to mask the target with his hand without pointing at it with the vision device. The duration of the whole second stage was recorded by the operator. The subject was given a 5 mn break between the two sessions. The last stage was the test. Forty targets were presented on the screen at 20 different positions on the subject's visual field (each one was presented twice). The subject, seated in front of the screen, was requested to "place the target at the center of his perceptive field, as accurately and quickly as possible". He had to validate when he felt it was the case. Position of the target in the visual space and time spent for each trial were recorded.

B. Durette, L. Gamond, S. Hanneton, D. Alleysson, J. Hérault

## 5. Results

Fourteen subjects, most of them students (m = 26 y, σ = 4 y) took part in the experiment; none of them had ever used this type of device. Seven were in the space-variant condition, seven in the uniform condition. Groups were paired for age and for gender. Results are described in fig 5.

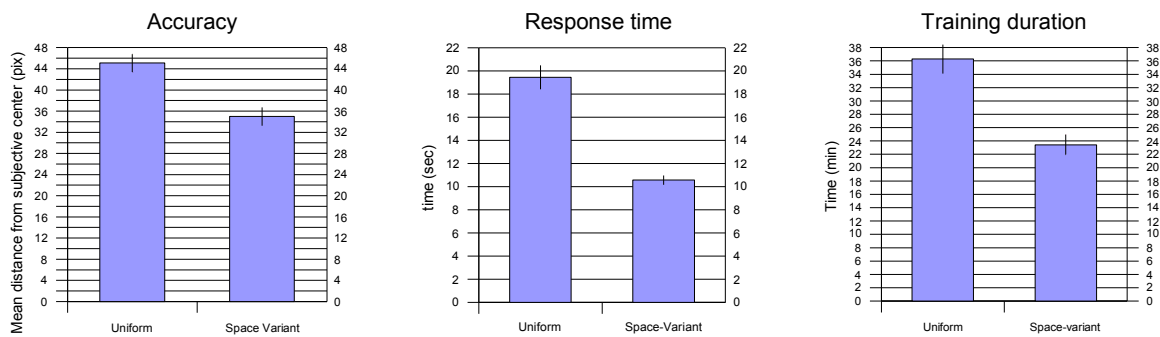

Figure 5: Performances in the localisation task for uniform and space-variant group. For all three measures, effect of the sampling distribution was highly significant (error bars: standard error).

*Target localization*: we computed the mean position for all trials, which we called the "subjective center". We then computed the distance between this subjective center and the actual position of the target at the end of each trial which is the data we used for assessing accuracy. Our hypothesis was that the accuracy should be enhanced when using a space-variant sampling. A unilateral Mann & Whitney test was applied to address this question. Subjects in space-variant condition (Log2) performed significantly better ($z = 4,99$ ; $p_{uni} < 0.001$) with mean $m_{log2}$ = 35 pix. and standard deviation $\sigma_{log2}$ = 29, against $m_{unif}$ = 44 pix., $\sigma_{unif}$ = 28 in the uniform condition.

*Time to focus the target*: the mean time required to focus at a visual target in test stage has been measured to test the efficiency of the device. Subjects in space-variant condition performed significantly better (Unilateral Mann & Whitney test, $z = 6,96$ ; $p_{uni} < 0.001$) with mean $m_{log2}$ = 10,58 s and standard deviation $\sigma_{log2}$ = 6,68 s, against $m_{unif}$ = 19 ,46 s, $\sigma_{unif}$ = 17,47s in the uniform condition.

*Training duration*: the duration of the second stage of training was also measured in order to assess difficulty in the device appropriation. As we had no idea about the direction of this effect, we performed a bilateral Mann & Whitney test. Subjects in space-variant condition performed significantly better (U= 47,5; $p_{bil} < 0.01$) with mean $m_{log2}$ = 23,4 min and standard deviation $\sigma_{log2}$ = 4,6min, against $m_{unif}$ = 36,3 min, $\sigma_{unif}$ = 6,7 min in the uniform condition.

Thus, with a shorter learning time and shorter response time, subjects in the space variant condition performed significantly better in terms of accuracy.

## 6. Discussion

Although quite promising, these results need to be carefully considered, for two primary reasons. When applied to a rectangular image, our mapping induces blank zones that are not sampled (at the borders). The proportion of this effect is related to the magnification factor. For R0=2, blank zones are approximately 10% of the picture. Although this aspect is not likely to explain our results, its effect can not be excluded. We are currently designing circular retinas that will not endorse this effect. Another possible caveat is the fact that the operator was aware of the subject's group when conducting the experiment. Even though the operator could not have had direct influence on the measures, results need to be confirmed with a double blind protocol.

However, this study shows that new pathways exist which can possibly enhance vision prosthesis, even though their resolution is bound to be limited. It provides an experimental framework to address their practical performances and it shows that significant and even counter-intuitive effects may be obtained. The fact that accuracy would be enhanced was natural since the sampling network was denser at the foveal center. Shorter response time may also be understood as an effect of the sampling variation, which provided additional information to locate a position in the visual field. On the other hand, space-variant sampling could have complicated the mastering of the vision device: it appears to be the opposite. This result leads to the idea that our mapping may facilitate the mastering of sensori-motor laws. This last aspect needs to be addressed further to determine whether our mapping is optimal or whether other kinds of space-variant sampling, a linearly decreasing distribution law for instance, may have the same effect.

Lastly, one may take a practical view of this study. To date, vision devices are far from giving blind people new eyes. However, contrary to surgical approaches (cataract operations, retinal transplant), electronic vision devices may easily be adapted to specific tasks. By giving the subject a better ability to locate a visual target, space variant sampling may be of use in devices designed for orientation, object finding and mobility assistance.

**Acknowledgments:** The "Sensory Substitution System for Visual Handicap" project is granted by the Rhône-Alpes region and the association "Les Gueules Cassées". Many thanks to K. O'Regan, S. Chokron, M. Auvray and C. Schoonover for fruitful discussions, and to C. Costello for linguistic support.